# ELECTRET-BASED CANTILEVER ENERGY HARVESTER: DESIGN AND OPTIMIZATION

**Sebastien Boisseau[1]\*, G. Despesse[1], A. Sylvestre[2]**
[1]CEA, LETI, Minatec, Grenoble, France
[2]G2ELab, Grenoble, France
*Presenting Author: s.boisseau@gmail.com

**Abstract:** We report in this paper the design, the optimization and the fabrication of an electret-based cantilever energy harvester. We develop the mechanical and the electrostatic equations of such a device and its implementation using Finite Elements (FEM) and Matlab in order to get an accurate model. This model is then used in an optimization process. A macroscopic prototype (3.2cm²) was built with a silicon cantilever and a Teflon® electret. It enabled us to harvest 17µW with ambient-type vibrations of 0.2g on a load of 210MΩ. The experimental results are in agreement with simulation results.

**Keywords:** electrets, vibration energy harvesting, electrostatic converters, Finite Element Model, optimization

## INTRODUCTION

Electret-based energy harvesters are known since the 1970s [1] but their spreading out rose only in the early 2000s with the development of MEMS. The goal of these devices is to harvest energy from vibrations, that is to say, to convert mechanical energy into electricity. Electret-based energy harvesters are part of electrostatic energy harvesters: therefore, they are based on a capacitive structure made of an electrode and a counter-electrode spaced by an air gap and polarized by an electret (a stable electrically charged dielectric). Vibrations induce changes in the capacitance geometry and a charges circulation between electrodes when they are connected by an electric circuit.

In this paper, we firstly describe the mechanical and the electrostatic equations of an electret-based cantilever energy harvester. Then, we present our model of this structure part using FEM (Finite Element Method) to compute the exact value of the capacitance and Matlab. An optimization process is then developed to maximize the output power. Finally, we present our experimental results obtained with a macroscopic prototype.

## HARVESTER DESCRIPTION – THEORY
### Energy harvester

Our electret-based cantilever energy harvester is made of a clamped-free conductive beam placed above an electret and an electrode and spaced by an air gap (Fig. 1). A mass is added at the free end of the cantilever. The conductive beam and the electrode are connected by a load ($R$).

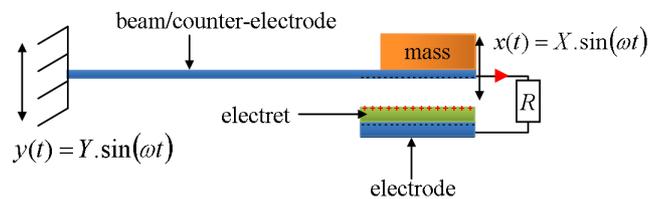

*Fig. 1: Model of the energy harvester.*

The study of the energy harvester can be separated into two parts: a study of the mechanical structure and a study of the electrostatic converter.

### Mechanical structure

The energy harvester can be modeled as a mobile mass ($m$) maintained in a support by a spring ($k$) and damped by forces ($f_{elec}$ and $f_{mec}$) [2]. When a vibration occurs $y(t) = Y\sin(\omega t)$, it induces a relative displacement of the mobile mass $x(t) = X\sin(\omega t + \varphi)$ compared to the frame (Fig. 2). A part of the kinetic energy of the mass is converted into electricity (modeled by an electrostatic force $f_{elec}$), while an other part is lost in friction forces (modeled by $f_{mec}$).

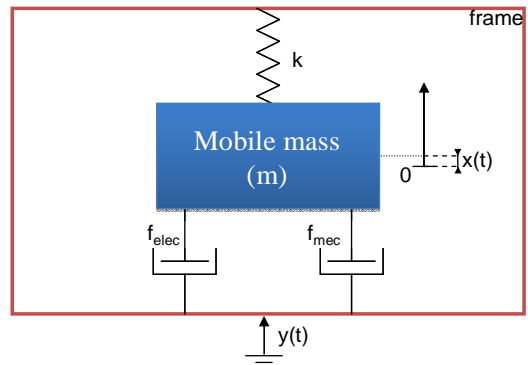

*Fig. 2: Equivalent model of the mechanical structure.*

This mass-spring structure is resonant and enables to take advantage of resonance phenomenon that amplifies the displacement amplitude of the mobile mass.

## Electrostatic converter

The goal of the electrostatic converter is to turn a mechanical relative displacement into electricity. The converter is made of a counter-electrode and an electrode on which is deposited an electret, spaced by an air gap and connected by a resistor (Fig. 3). The electret has a constant charge $Q_i$, and, due to electrostatic induction and charges conservation, the sum of charges on the electrode and on the counter-electrode equals the charge on the electret: $Q_i = Q_1 + Q_2$.

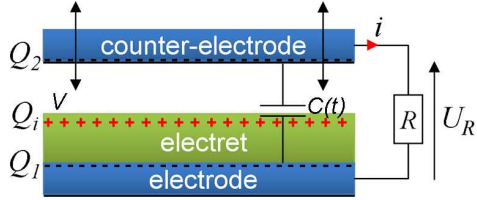

Fig. 3: Model of the electrostatic converter.

When a vibration occurs, it induces a change in the geometry of the capacitor $C$ (the counter-electrodes moves away from the electret, changing the average air gap) and a reorganization of charges between the electrode and the counter-electrode through the load. This induces a current across the load $R$ and mechanical energy is turned into electricity.

## Whole structure and equations

By applying Newton's law to the mechanical structure and Kirchhoff's law to the electrostatic converter, it can be demonstrated that the electret-based cantilever energy harvester is constrained by the following equation system:

$$\begin{cases} m\ddot{x} + b_m \dot{x} + kx - \dfrac{d}{dx}\left(\dfrac{Q_2^2}{2C(t)}\right) - mg = -m\ddot{y} \\ \dfrac{dQ_2}{dt} = \dfrac{V}{R} - \dfrac{Q_2}{C(t)R} \end{cases} \quad (1)$$

Where $V$ is the surface voltage of the electret.

Cantilever energy harvesters are quite simple but present the advantage of being particularly efficient when working with constant vibrations. Actually, since these structures are resonant, they are interesting only if their natural frequency is tuned to the frequency of the ambient vibrations. Moreover, they avoid the challenge of electret patterning that can lead to a stability decrease and a cost increase of the electret. Fig. 4 presents the design of the energy harvester we want to make.

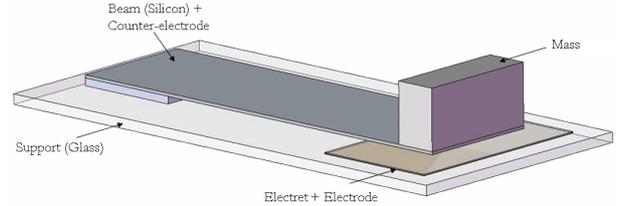

Fig. 4: Schema of the future prototype.

To extract the maximum energy from the environment, the parameters of the device have to be optimized. It is first necessary to develop a precise model of the structure.

# FINITE ELEMENTS AND OPTIMIZATION PROCESS

In order to get an accurate model of the energy harvester, we have decided to use finite elements.

## Finite Elements Model

The goal of finite elements is to compute the capacitance $C$ between the electrode and the beam for a given displacement ($x$). Actually, an other possibility could be to model the capacitor as a plane capacitor but, it induces important mistakes on the capacitance value because of the deformation shape of the beam and fringe effects.

To develop the finite element method on this structure, we have chosen Comsol® Multiphysics. Our FEM model uses plane stress (*ps*), moving mesh (*ALE*) and electrostatic (*es*) modules of Comsol Multiphysics. The capacitance is computed by applying a voltage to the beam ($U_b$) and by putting the electrode to *GND* for a given displacement at the free end of the beam (Fig. 5).

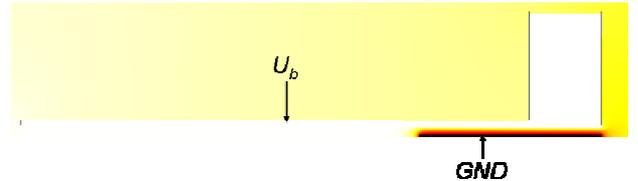

Fig. 5: FEM model - Potential.

Comsol® can compute the electrostatic energy ($W_e$) of the system. Then, it is easy to get the capacitance ($C$) from the electrostatic energy ($W_e$) since:

$$W_e = \frac{1}{2}CU_b^2 \quad (2)$$

## Optimization Process

It is not possible to solve Eq. 1 by calculation and to get an analytic expression of the output power of the energy harvester (nonlinear coupled differential equations). Nevertheless, it is possible to solve this problem numerically. This has been achieved with Matlab/Simulink (Fig. 6). Eq. 1 is represented as blocks and the "capacitance block" directly uses

Comsol to compute the capacitance.

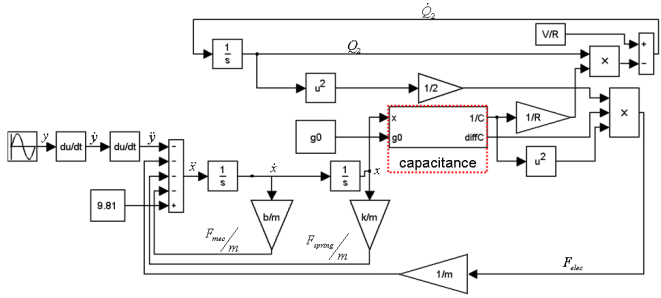

*Fig. 6: Simulink model of the energy harvester.*

Thanks to this model, it is possible to develop an optimization process using the *fminsearch* Matlab function. The parameter to optimize is the average output power ($P$) of the energy harvester, with:

$$P = \frac{1}{t_2 - t_1} \int_{t_1}^{t_2} R \left( \frac{dQ_2}{dt} \right)^2 dt \quad (3)$$

Where $t_1$ and $t_2$ are times taken in the steady state.

The goal of the optimization process is to find the best values for the load ($R$), the initial air gap ($g_0$) and the surface of the electrodes ($S$) for a given vibration ($Y$, $\omega$) and a given surface voltage ($V$). Actually, these parameters have an important effect on the output power. Fig. 7 presents for example the effect of the load on the output power and proves clearly the existence of an optimum for this parameter.

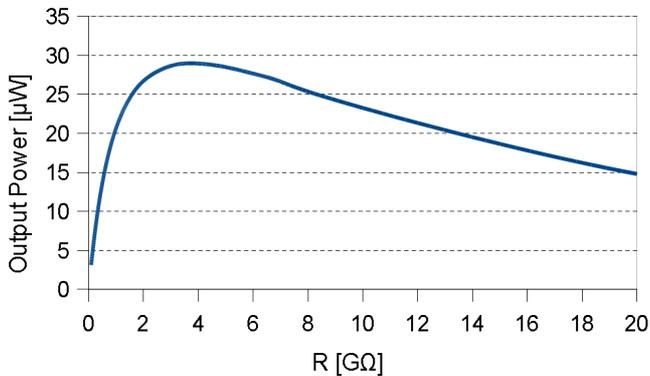

*Fig. 7: Effect of the load (example).*

Our beam is made in silicon, sizes 3.2cm×9mm×300µm and has a natural frequency of 45Hz with a mobile mass of 3g. The optimization process gave the following optimized values (Table 1) for our beam and for a vibration of 46µm$_{peak-to-peak}$@45Hz.

*Table 1: Optimized parameters.*

| Parameter | Optimized value |
|---|---|
| $R$ | 210MΩ |
| $g_0$ | 2.2mm |
| $S$ | 8mm×9mm |

The output power should be 30µW and the expected output voltage is presented in Fig. 8.

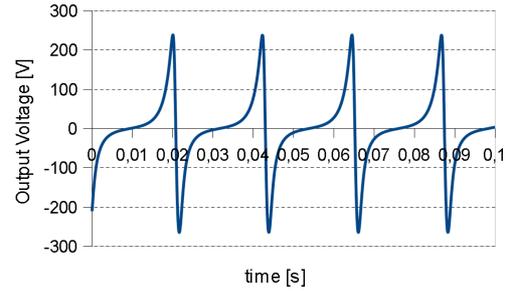

*Fig. 8: Expected output voltage.*

The output voltages of the energy harvester are high; this should simplify a future AC/DC conversion.

**EXPERIMENTAL RESULTS**
**Prototype manufacturing**
The optimal parameters were applied to a prototype based on our silicon beam and a Teflon® electret (Fig. 9). The mechanical quality factor of the beam has been measured and equals 75. The electret is 100µm-thick, charged to 1200V and has a surface of 0.72cm². The stability of the electret was not studied in this work. Nevertheless, no important charge decay was observed during the some days of the experiments.

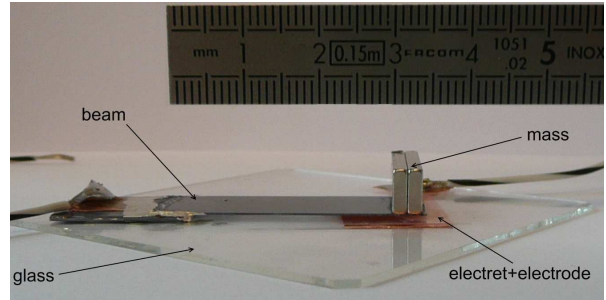

*Fig. 9: Energy harvester – Prototype.*

**Experimental output power**
With 'ambient' vibrations of 46µm$_{peak-to-peak}$@45Hz (~0.2G), 17µW are experimentally harvested on a load of 210MΩ. The figure below that represents the output voltage proves that simulation curves are in agreement with experimental curves (Fig. 10).

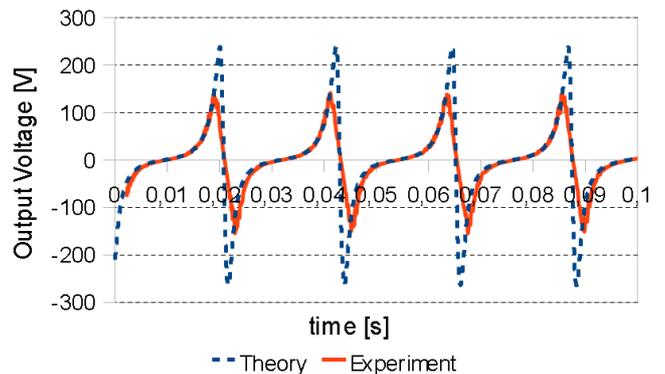

*Fig. 10: Experimental output voltage for a load of 210MΩ.*

*Table 2: Most recent electret energy harvesters in the state of the art.*

| Author | Ref | Vibrations (amplitude peak-to-peak: $2Y$, frequency: $f$)[1] | Active Surface ($S$) | Electret Potential ($V$) | Output Power ($P$) | Figure of merit $\chi$ $\chi = \dfrac{P}{Y^2(2\pi f)^3 S}$ |
|---|---|---|---|---|---|---|
| Suzuki | [3] | 2mm$_{pp}$@37Hz | 2.33 cm² | 450V | 0.28 µW | 9.56×10$^{-5}$ |
| Sakane | [4] | 1.2mm$_{pp}$@20Hz | 4 cm² | 640V | 0.7 mW | 2.45 |
| Naruse | [5] | 50mm$_{pp}$@2Hz | 9 cm² |  | 40µW | 3.58×10$^{-2}$ |
| Halvorsen | [6] | 5.6µm$_{pp}$@596Hz | 0.48 cm² |  | 1µW | 5.06×10$^{-2}$ |
| Kloub | [7] | 0.16µm$_{pp}$@1740Hz | 0.42 cm² | 25V | 5µW | 14.2 |
| Edamoto | [8] | 1mm$_{pp}$@21Hz | 3 cm² | 600 V | 12µW | 6.97×10$^{-2}$ |
| Miki | [9] | 0.2mm$_{pp}$@63Hz | 3 cm² | 180V | 1µW | 5.37×10$^{-3}$ |
| Honzumi | [10] | 18.7µm$_{pp}$@500Hz | 0.01 cm² | 52V | 90 pW | 3.32×10$^{-5}$ |
| This work |  | 46µm$_{pp}$@45Hz | 3.2 cm² | 1200V | 17 µW | 4.44 |

[1] pp stands for peak-to-peak

## DISCUSSION

These experimental results are compared to the state of the art in Table 2 and prove that, with a simple structure, it is possible to reach the magnitude of harvested powers in the state of the art.

Therefore, the simple structure of the electret-based cantilever energy harvester can be a good solution when the vibrations are constant in terms of frequency and amplitude because the initial gap can be optimized. Nevertheless, when the vibrations are not constant, these structures are not a good solution: if the natural frequency is not tuned to the vibration frequency, the relative displacement of the beam compared to the electrode is low and the output power will be limited. Moreover, if the vibration amplitude increases, it induces risks of collisions between the beam and the electret and may reduce the life of the electret. Similarly, if the vibration amplitude decreases, the capacitance variation is not large enough and low output power is generated.

## CONCLUSION

We demonstrate that electret-based cantilever energy harvester can be a good solution to harvest energy from vibrations when the vibrations are constant in frequency and amplitude. Thus, thanks to a good optimization, the output powers are similar to the one gotten in the state of the art. Of course, if the vibrations are not constant these structures lose their interest and other structures such as electret patterned in-plane energy harvesters or piezoelectric cantilevers should be considered.

## REFERENCES

[1] Jefimenko O D 1978 Electrostatic current generator having a disk electret as an active element *IEEE Trans. Indust. Appl.* **14** 537–540

[2] Williams C B, Yates R B 1996 Analysis of a micro electric generator for microsystems *Sensors Actuators A* **52** 8–11

[3] Suzuki Y, Edamoto M, Kasagi N, Kashwagi K, Morizawa Y 2008 Micro electret energy harvesting device with analogue impedance conversion circuit *Proc. PowerMEMS'08* 7–10

[4] Sakane Y et al 2008 The development of a high-performance perfluorinated polymer electret and its application to micro power generation *J. Micromech. Microeng.* **18** 104011

[5] Naruse Y, Matsubara N, Mabuchi K, Izumi M, Suzuki S 2009 Electrostatic micro power generation from low-frequency vibration such as human motion *J. Micromech. Microeng.* **19** 094002

[6] Halvorsen E, Westby E R, Husa S, Vogl A, Østbø N P, Leonov V, Sterken T, Kvisterøy T 2009 An electrostatic energy harvester with electret bias *Proc. Transducers'09* 1381–1384

[7] Kloub H, Hoffmann D, Folkmer B, Manoli Y 2009 A micro capacitive vibration energy harvester for low power electronics *Proc. PowerMEMS'09* 165–168

[8] Edamoto M et al 2009 Low-resonant-frequency micro electret generator for energy harvesting application *Proc. MEMS'09* 1059–1062

[9] Miki D, Honzumi M, Suzuki Y, Kasagi N 2010 Large-amplitude MEMS electret generator with nonlinear spring *Proc. MEMS'2010* 176–179

[10] Honzumi M et al 2010 Soft-x-ray-charged vertical electrets and its application to electrostatic transducers *Proc. MEMS'2010* 635–638